\documentclass[10pt,a4paper]{spie}
\usepackage[utf8]{inputenc}
\usepackage[dvipdf]{graphicx}
\usepackage{makeidx}
\usepackage{color}
\usepackage{multicol}

\title{
    Investigating the use of the hydrogen cyanide (HCN) as an absorption media for laser spectroscopy
    }

\author{
	Martin Hosek,
    Simon Rerucha,
	Lenka Pravdova,
	Martin Cizek,\\
	Jan Hrabina,
	Petr Jedlicka and
	Ondrej Cip 
\skiplinehalf
Institute of Scientific Instruments of the Czech Academy of Sciences (ISI),\\
 Kr\'{a}lovopolsk\'{a} 147, 612 64 Brno, Czech Republic
\skiplinehalf
}

\authorinfo{Further author information: please send correspondence to e-mail address mhosek@isibrno.cz (M. Hosek)}

\begin{document}
\maketitle

%%%%%%%%%%%%%%%%%%%%%%%%%%%%%%%%%%%%%%%%%%%%%%%
% hlavicka textu
%%%%%%%%%%%%%%%%%%%%%%%%%%%%%%%%%%%%%%%%%%%%%%%

\begin{abstract}
The laser spectroscopy is a fundamental approach for the realisation of traceable optical frequency standards.
In the $1.55\,\mu$m wavelength band, widely used in telecommunications, the acetylene is the typical and the most
widespread absorption media. We present our investigation of using the hydrogen cyanide (HCN) as a cost-
efficient and readily available alternative, that also provides a wider frequency span (from $1527\,$nm to $1563\,$nm).
We have compared the practical aspects of using new absorption media in comparison to existing experience
with the acetylene with an outlook to carry out an independent measurement of the entire spectra. The results
should contribute to the future inclusion of the HCN spectroscopic data into the Mise en pratique, thus allowing
for the use of HCN as a reference for the realisation of traceable laser standards.
\end{abstract}

\\
\\
{\bf DOI:} 10.1117/12.2517761 \\

%
% 
%
%

%%%%%%%%%%%%%%%%%%%%%%%%%%%%%%%%%%%%%%%%%%%%%%%
% uvod
%%%%%%%%%%%%%%%%%%%%%%%%%%%%%%%%%%%%%%%%%%%%%%%

\section{Introduction}

Even though the He-Ne lasers remains the main source of optical frequency for for the metrology of length
(lasing close to the wavelength of $633\,$nm), the popularity of the $1.55\,\nu$m is also increasing. The advantage
of this band over the 633nm band is that the optical components are easier for production, cheaper and the
technological possibilities for the laser sources and optical components are more advanced.

The most widely used spectroscopic media for building the frequency standards\cite{onae1999} for the $1.55\,\nu$m is the
acetylene ($^{13}C_2H_2$ and $^{12}C_2H_2$ isotopes) with its absorption lines in this band well described\cite{edwards2005} with the associated
uncertainty in order of $10^{11}$. 
The alternative wavelength reference is the hydrogen cyanide (HCN), which is
better available then acetylene (especially the 13C2H2 isotope). The spectrum of HCN is also richer with wider
frequency span, which could lead to development of the laser-based optical frequency standards with a wider
range of optical frequencies.
The analysis of the HCN absorption lines has been already done, but due to technological limitations at the
time of measurement they are not as precise as the data for acetylene (uncertainty in order of $10^{6}$)\cite{gilbert1998}.
Our approach features several improvements specically the lock-in of the laser frequency on the particular
transition and the direct measurement of the beat-note between the laser and the optical frequency comb (OFC).

%%%%%%%%%%%%%%%%%%%%%%%%%%%%%%%%%%%%%%%%%%%%%%%
% background a metudy
%%%%%%%%%%%%%%%%%%%%%%%%%%%%%%%%%%%%%%%%%%%%%%%

\section{Methods}
\label{met}

For the first analysis of $H^{13}C^{14}N$ cell (made by Wavelength References, Inc., $20\,$cm length, nominal pressure$67\,$Pa) we used linear absorption spectroscopy\cite{kreuzer1971}. Its advantage is the relative simplicity of experimental set-up, which making it ideal method for the initial measurement. For the initial assembling of the experimental set-up, the $^{13}C_2H_2$ cell (made at ISI\cite{hrabina2014,lazar2009}, $30\,$cm length, $100\,$Pa nominal pressure) was used as $^{13}C_2H_2$ is a well know absorption media and it helped us to set-up the experimental arrangement with the a-priori experience.

\begin{figure}[htbp]
	\centering
	\includegraphics[width=.8\textwidth]{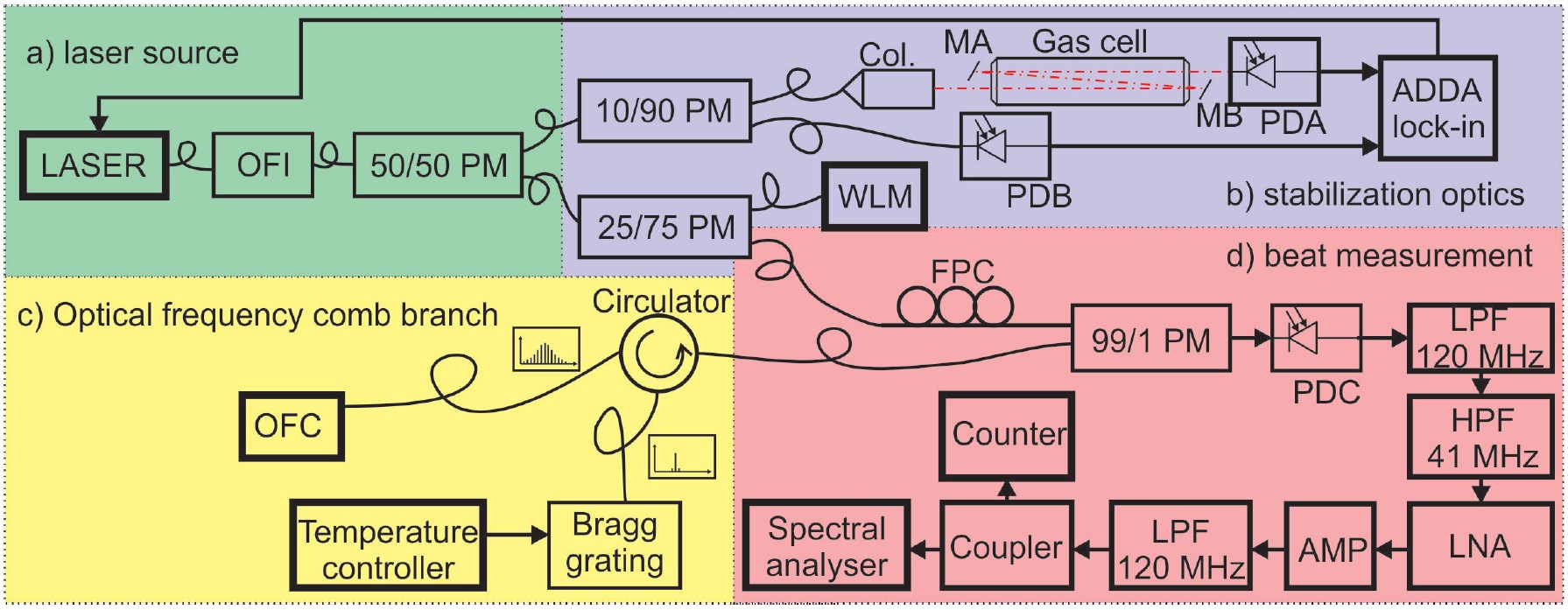}
	
	\caption{Scheme of experimental set-up. The laser source part contains the laser, optical frequency isolator (OFI) and 50/50 polarization maintaining (50/50 PM) beam-splitter. The stabilization optics part consists of 25/75 polarization maintaining (25/75 PM) beam splitter, 10/90 polarization maintaining (10/90 PM) beam splitter, collimator (Col.), gas cell, two mirrors (MA and MB), two phodotodetectors (PDA and PDB), analog-digital lock-in (ADDA lock-in) and wavelength-meter. The optical frequency comb branch comprise of optical frequency comb (OFC), circulator, temperature controller and Bragg grating. The beat measurement part includes fibre polarization controller (FPC), 99/1 polarization maintaining (99/1 PM) beam-splitter, photodetector (PDC), two low pass filters with the upper limit of 120MHz (LPF 120 MHz), high pass filter with the lower limit of 41MHz (HPF 41 MHz), low noise amplifier (LNA), amplifier (AMP), coupler, frequency counter and spectral analyser}
	\label{f1}
\end{figure}

\subsection{Experimental setup}

In our experimental setup, shown in Figure \ref{f1}, we used tunable Koheras Adjustik K-81 laser with tuning range approximately from $1539.8\,$nm to $1541.0\,$nm. The laser beam passed through the optical Faraday isolator (OFI), which served for preventing the reflected light to enter the laser. The beam was split 50/50, where one part propagated into the stabilization part of the set-up and the second into the part for measurement of the beat-note of laser with the optical frequency comb (OFC).

The laser beam in the stabilization part of the experimental set-up passed through the 90/10 beam splitter
where the less intensive part of the beam was sent directly to the photodetector (PDB). The more intensive part
of the laser beam passed through the collimator and then passed through the gas cell three times. After that
the beam finally entered the photodetector (PDA). During the tuning of the laser, their frequency (wavelength)
was measured by the wavelength meter (WLM).

The beam for the beat-note measurement passed through the fibre polarization controller (FPC), which
served for the control of the polarization of the beam. The beam from the laser was then combined with the
beam from OFC.

The signal from OFC passed through the circulator and hit the Bragg grating used for the separation of the
part of OFC spectrum, which frequency was similar to the frequency of the laser. The temperature controller
was used for the control of the temperature of Bragg grating, but as the temperature dependence of re
ected
light from the grating was negligible the temperature was left the same during the whole experiment ($18.992\,^o$C).

The signal of the OFC and laser were combined by beam splitter 99/1 and hit the photodetector (PD). Then
the signal passed through the combination of high pass filter (HPF), low pass filters (LPF) , low noise amplifier
(LNA) and amplifier (AMP) and the frequency of the beat-note was measured by the counter.

\subsection{Measurement procedure}

\begin{figure}[htbp]
	\centering
	\includegraphics[width=.5\textwidth]{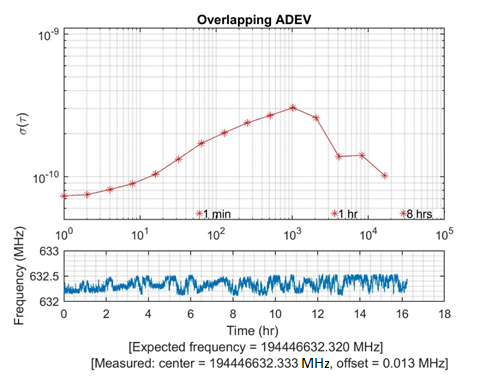}
	
	\caption{a) overlapping Allan deviation demonstrating the achievable stability of the frequency of laser being locked by our lock-in system. b) The time dependence of the laser frequency}
	\label{f2}
\end{figure}

At first, for both measured gases ( $^{13}C_2H_2$ and  $^{13}C_2H_2$) we measured the spectrum of the gas by fast tuning of the laser across the available frequency range. This allowed us to determine which absorption lines are involved in our spectra and to identify them. After that we measured the profiles of the identified lines by slow tuning over them. The profiles of the lines were fitted by Voigt profile and the linewidth of each line was calculated. Finally we locked the laser on the center of the corresponding line and measured the stability of the laser and also determined the position of the center of the line by the beat-note measurement.
The laser frequency was locked on the center of the absorption lines by in-house custom-build lock-in hardware.
The initial purpose of the software was to guarantee the long-term stability of laser diodes. The achievable
stability by this lock-in hardware is shown in Figure \ref{f2}.

We used beat-note between laser and optical frequency comb\cite{cizek2014} for more precise determination of the laser beam frequency, when the laser was locked to the center of absorption line. The Bragg grating served for the separation of the proper part of the OFC spectrum. The frequency of the laser $\nu_{laser}$ was calculated from the frequency of the beat-note $\nu_{beat}$ by the following equation

\begin{equation}
\nu_{laser} = n \cdot \nu_{rep} + \nu_{off} \pm \nu_{beat}
\end{equation}

where $n$ is the order of the OFC, $\nu_{rep}$ is the repetition frequency of the OFC and $\nu_{off}$ is the offset frequency of the OFC.
The uncertainty of the results was calculated with the coverage factor $k = 2$, which means that the true value
of the measured quantity was in the confidence interval with $95.\,5$\% probability.

%%%%%%%%%%%%%%%%%%%%%%%%%%%%%%%%%%%%%%%%%%%%%%%
% resultety
%%%%%%%%%%%%%%%%%%%%%%%%%%%%%%%%%%%%%%%%%%%%%%%

\section{Results}
\label{s3}
\subsection{Acetylene}

\begin{figure}[htbp]
	\centering
	\includegraphics[width=.8\textwidth]{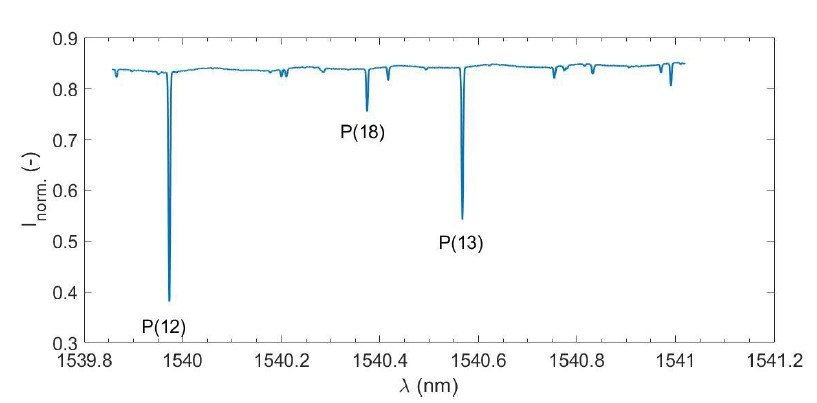}
	
	\caption{The spectrum of $^{13}C_2H_2$ measured in the range from 1539.8 nm to 1541 nm.}
	\label{f3}
\end{figure}

At first, we analysed the  $^{13}C_2H_2$ to determine the properties of our experimental set-up. We quickly tuned the laser over the available spectral range and recorded the signal from the photodetectors. 
We divided the signal behind the cell by the signal in front of it to minimize the in
uence of 
uctuation of power of the laser on the measured absorption spectrum. In the spectrum (see Figure \ref{f3}.), we were able to recognize three strong lines of acetylene. We measured the profile of each line by slow frequency detuning across the spectral line, then we fitted The line profiles by Voigt profile and determined the linewidth of each line (Table \ref{table1}).

 \begin{table}[!t]
	\vskip 2mm
	\renewcommand{\arraystretch}{1.3}
	\caption{The results obtained for the absorption lines of $^{13}C_2H_2$. $\nu_{peak}$ stands for the frequency of the center of absorption
		line, $\nu_{lw}$ stands for the linewidth of absorption line, $\nu_{ref}$ stands for the reference frequency of the center of absorption
		line and $\Delta\nu$ stands for $|\nu_{ref} - \nu_{peak}|$}
	\label{table1}
	\centering
	\begin{tabular}{l l c c }
		
		$\nu_{peak}$ (MHz) 		& $\nu_{ref}$\cite{gilbert1998} (MHz)	& $\nu_{lw}$ (MHz)	& $\Delta\nu$ (MHz)\\
		\hline
		194 673 775.1(35) 	&194 673 775.9104(9) 	&430 	&0.8104 \\
		194 623 106.7(226) &194 623 100.1112(26) 	&460 	&6.5888 \\
		194 598 734.8(5) 	&194 598 735.3505(5) 	&440 	&0.5505 \\
		\hline
	\end{tabular}
\end{table}

\begin{figure}[htbp]
	\centering
	\includegraphics[width=.6\textwidth]{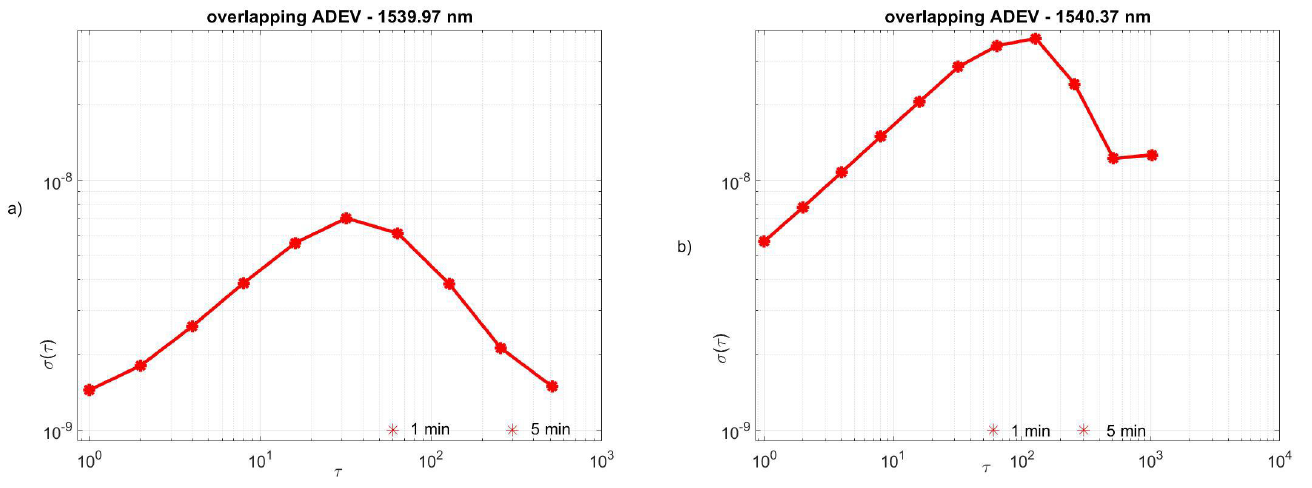}
	\includegraphics[width=.3\textwidth]{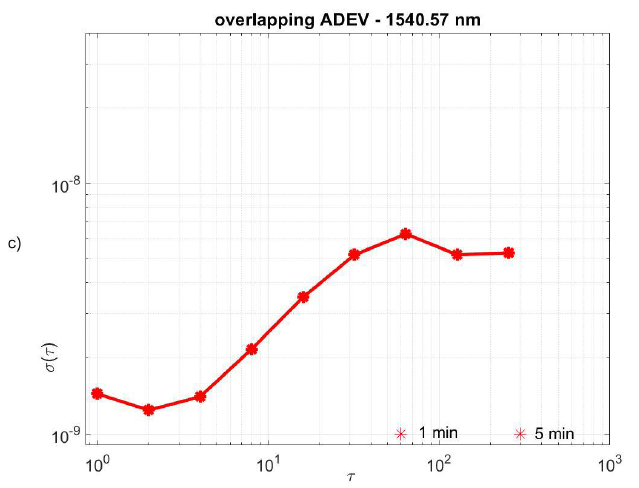}
	\caption{The overlapping Allan deviations calculated for the laser frequency being locked on the $^{13}C_2H_2$ absorption lines: a) P(12) line in $\nu1 + \nu2$ band, b) P(18) line $\nu1 + \nu2 + \nu4 + \nu5$ band, c) P(13) line in $\nu1 + \nu2$ band}
	\label{f4}
\end{figure}

After the identification of the absorption lines of $^{13}C_2H_2$ in the spectrum, the optical frequency of the laser was successively locked on the minimum of each of the absorption lines and kept in the lock about 10 minutes. The frequency of the center of the absorption lines was calculated by the measurement of beat-note of laser with the OFC. Observed center frequencies as well as the differences between them and the published values are shown in Table \ref{table1}. The stability of the laser was then estimated by calculating the overlapping Allan deviation (Figure \ref{f4}).

\subsection{Hydrogen cyanide}

\begin{figure}[htbp]
	\centering
	\includegraphics[width=.9\textwidth]{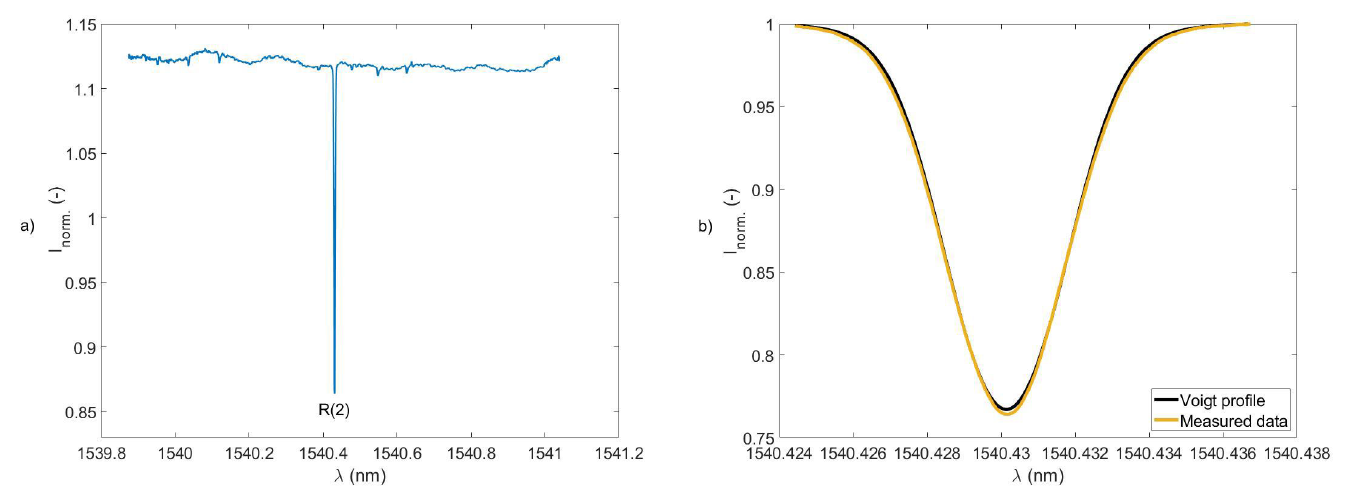}
	\caption{a) The spectrum of $H^{13}C^{14}N$ measured in the range from 1539.8nm to 1541 nm, b) The line profile of R(2) line of $H^{13}C^{14}N$}
	\label{f5}
\end{figure}

\begin{figure}[htbp]
	\centering
	\includegraphics[width=.35\textwidth]{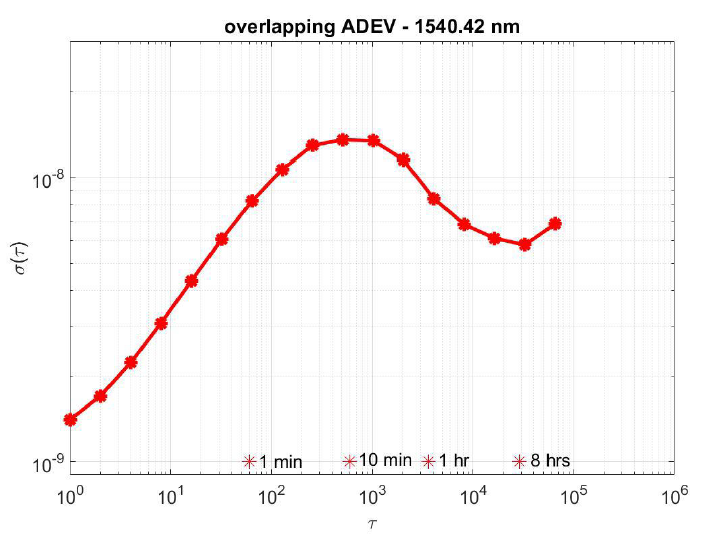}
	\caption{The overlapping Allan deviation calculated for the laser frequency being locked on the $H^{13}C^{14}N$ absorption line R(2) }
		\label{f6}
\end{figure}

After alignment of our system, the $H^{13}C^{14}N$ gas was analysed. We measured its absorption spectrum by the slow frequency detuning across the spectrum (Figure \ref{f5}. part b)). In the measured spectrum of $H^{13}C^{14}N$ was identified one strong absorption line. The line profile (Figure \ref{f5}b)) of this line measured by slow scanning across its profile and the calculated linewidth is shown in Table \ref{table2}. The laser was subsequently locked on the absorption line of $H^{13}C^{14}N$ and the stability of the laser was determined by the calculation of overlapping Allan deviation Figure \ref{f6}. The frequency of the center of the absorption line of $H^{13}C^{14}N$ was determined by the beat-note with the OFC and is shown in Table \ref{table2}. The difference between the measured value and published value is quite high, but the measured value is more precise and it is within the confidence interval of published value.

\begin{table}[!t]
	\vskip 2mm
	\renewcommand{\arraystretch}{1.3}
	\caption{
		The results obtained by fitting of the absorption line of $H^{13}C^{14}N$ and subsequently by lock-in of 	the laser. $\nu_{peak}$ stands for the frequency of the center of absorption line, $\nu_{lw}$ stands for the linewidth of absorption line, $\nu_{ref}$ stands for the reference frequency of the center of absorption line, $\Delta\nu_{ref}$ stands for $|\nu_{ref} - \nu_{peak}|$ and $\Delta\nu_{cal}$ stands for $|\nu_{cal} - \nu_{peak}|$. The $\nu_{ref}$ value was calculated from the published value\cite{gilbert1998} in wavelength and the $\nu_{cal}$ was calculated from theoretically estimated value\cite{swann2005} in wavelength.}	
	\label{table2}
	\centering
	\begin{tabular}{l l l c c c }
		
		$\nu_{peak}$ (MHz) 		& $\nu_{ref}$\cite{gilbert1998} (MHz) & $\nu_{cal}$\cite{swann2005} (MHz)	& $\nu_{lw}$ (MHz)	& $\Delta\nu_{ref}$ (MHz) & $\Delta\nu_{cal}$ (MHz)\\
		\hline
		194 615 893.1 (101)& 194 615 960.1 (7580) &194 615 892.2 (10) &470 &67.0 &0.9 \\
		\hline
	\end{tabular}
\end{table}

%%%%%%%%%%%%%%%%%%%%%%%%%%%%%%%%%%%%%%%%%%%%%%%
% diskuse
%%%%%%%%%%%%%%%%%%%%%%%%%%%%%%%%%%%%%%%%%%%%%%%

\section{Discussion}

\begin{figure}[htbp]
	\centering
	\includegraphics[width=.6\textwidth]{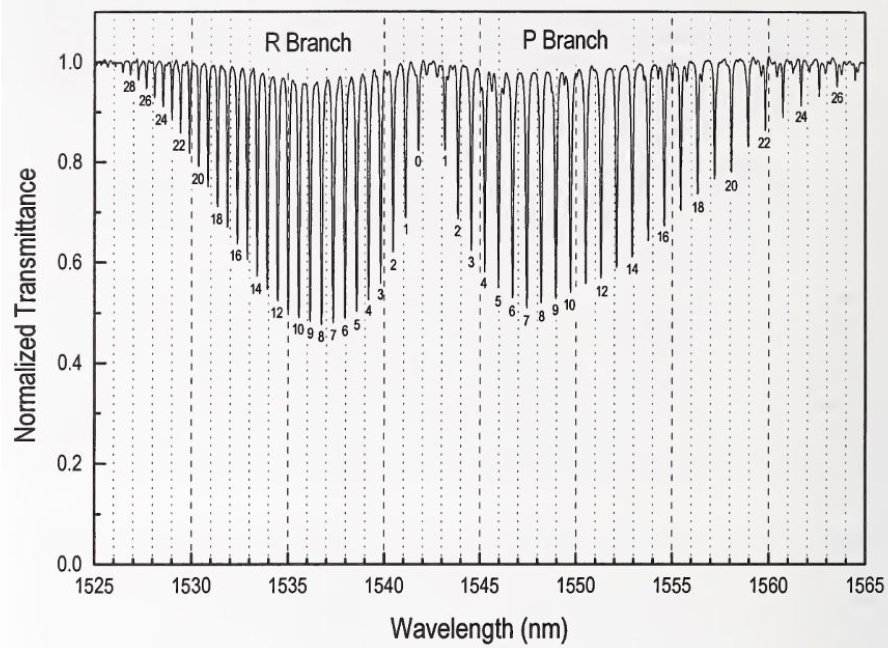}
	\caption{The absorption spectrum of $H^{13}C^{14}N$ in the spectral range from 1525 nm to 1565 nm}
		\label{f7}
	\end{figure}

We managed to got the possitions of the centers of absorption peak with the uncertainty of $10^{-8}$ for both
investigated gases (${13}C_2H_2$, $H^{13}C^{14}N$). Even though the presented results (Section \ref{s3}) are just very first results the precision of our results is in the case of $H^{13}C^{14}N$ better than the precision of results already published\cite{gilbert1998}.
From the calculated Allan deviations it is evident that the achievable stability is better by factor 10 to 100 then the stability we achieve in our measurement. Our primery goal will be to improve the stability to the $10^{-10}$ level and then to further improve the lock-in by use of saturation absorption spectroscopy to get even better stability close to $10^{-12}$ level. The saturation absorption spectroscopy\cite{lazar2009} will allow us to get rid of Doppler broadening of the absorption lines, which in case of our measurement was about $450\,$MHz. However, the preliminary tests with the $63\,$Pa cell revealed that the Doppler-free hyperfine transitions are not detectable. We measured just one absorption line of $H^{13}C^{14}N$ which was caused by too narrow tunability range of our laser. The previous measurements of the HCN spectrum indicate (see Figure \ref{f7}) that there are two absorption lines close to our spectral range, but not close enough to be actually measured. Our future goal is to do a survey in order to choose a suitable laser source, that will allow us to cover wider spectra and to measure more absorption lines of $H^{13}C^{14}N$.

%%%%%%%%%%%%%%%%%%%%%%%%%%%%%%%%%%%%%%%%%%%%%%%
% konkluze
%%%%%%%%%%%%%%%%%%%%%%%%%%%%%%%%%%%%%%%%%%%%%%%

\section{Conclusion}

We presented the first results of our work that aims at the improvement of the already accessible data for centres of absorption lines of $H^{13}C^{14}N$. We managed to obtain more precise data then are available even though there is still lot of space for improvement. The future goal will be to use widely-tunable laser, lock it to a hyperfine transition, and with the beat frequency measurement assembly, measure the exact frequency of the hyperfine transitions of the HCN for the wavelengths from 1527nm to 1563nm where the results will serve as material for a newly compiled atlas of the HCN transitions.

\acknowledgments 

The authors acknowledge the support from Academy of Sciences of the Czech Republic project RVO: 68081731
and Ministry of Education, Youth and Sports of the Czech Republic (LO1212) together with the European
Commission (ALISI No. CZ.1.05/2.1.00/01.0017); Parts of this research were performed within the EMPIR
project 17IND03 LaVA. The financial support of the EMPIR initiative is gratefully acknowledged. The EMPIR
initiative is co-funded by the European Union's Horizon 2020 research and innovation programme and EMPIR
Participating States. Particular issues has been supported by the Ministry of Industry and Trade of the Czech
Republic (project FV10336), and Technology Agency of the Czech Republic (project TE01020233).

%%%%%%%%%%%%%%%%%%%%%%%%%%%%%%%%%%%%%%%%%%%%%%%
% bibliografie
%%%%%%%%%%%%%%%%%%%%%%%%%%%%%%%%%%%%%%%%%%%%%%%

\end{document}